\documentclass[aps,prb,showpacs,twocolumn]{revtex4}
\usepackage{amsmath,graphics}
\usepackage[next]{inputenc}
\usepackage[dvips]{epsfig}
\def\be{\begin{equation}}
\def\ee{\end{equation}}

\def\bi{\begin{itemize}}
\def\ei{\end{itemize}}
\def\bn{\begin{enumerate}}
\def\en{\end{enumerate}}
\def\bea{\begin{eqnarray}}
\def\eea{\end{eqnarray}}
\def\no{\nonumber}
\def\ba{\begin{array}}
\def\ea{\end{array}}
\def\bd{\begin{displaymath}}
\def\ed{\end{displaymath}}

\begin{document}
\title{Entanglement properties of topological color codes}

\author{M. Kargarian}
\email[]{kargarian@physics.sharif.edu}

\affiliation{Physics Department, Sharif University of Technology,
Tehran 11155-9161, Iran}

\begin{abstract}
The entanglement properties of a class of topological stabilizer
states, the so called  \emph{topological color codes} defined on a
two-dimensional lattice or \emph{2-colex}, are calculated. The
topological entropy is used to measure the entanglement of different
bipartitions of the 2-colex. The dependency of the ground state
degeneracy on the genus of the surface shows that the color code can
support a topological order, and the contribution of the color in
its structure makes it interesting to compare with the Kitaev's
toric code. While a qubit is maximally entangled with rest of the
system, two qubits are no longer entangled showing that the color
code is genuinely multipartite entangled. For a convex region, it is
found that entanglement entropy depends only on the degrees of
freedom living on the boundary of two subsystems. The boundary
scaling of entropy is supplemented with a topological subleading
term which for a color code defined on a compact surface is twice
than the toric code. From the entanglement entropy we construct a
set of bipartitions in which the diverging term arising from the
boundary term is washed out, and the remaining non-vanishing term
will have a topological nature. Besides the color code on the
compact surface, we also analyze the entanglement properties of a
version of color code with border, i.e \emph{triangular color code}.

\end{abstract}
\date{\today}


\maketitle
%%%%%%%%%%%%%%%%%%%%%%%%%%%%%%%%%%%%%%%%%%%%%%%%%%%%%%%%%%%%%%%%%%%%%
\section{Introduction \label{introduction}}
In quantum information theory and quantum computations, entanglement
is recognized as an essential resource for quantum processing and
quantum communications, and it is believed that the protocols based
on the entangled states have an exponential speed-up than the
classical ones. Besides, in highly correlated states in condensed
matter systems such as superconductors\cite{oh,vedral}, fractional
quantum Hall liquids \cite{wen1}, the entanglement serves as a
unique measure of quantum correlations between degrees of freedom.
In quantum many-body systems such as spin, fermion and boson
systems,the entanglement is connected to the phase diagram of the
physical systems. Non-analytic behavior of entanglement close to the
quantum critical point of the system and occurrence of finite size
scaling have provided an intense research leading to fresh our
insight of the critical properties from the quantum information
side, for a comprehensive discussion see the review by L. Amico,
\emph{et al}. \cite{amico} and references therein. In the past years
the appearance of new phases of matter has intensified the
investigation of the entanglement in the quantum systems. These are
phases beyond the Landau-Ginzburg-Wilson paradigm \cite{goldenfeld}
where an appropriate local order parameter characterizes different
behaviors of two phases on either side of the critical point. These
new phases of matter carries a kind of quantum order called
\emph{topological order}\cite{wen2} and the transition among various
phases does not depend on the symmetry breaking mechanism. Therefore
the Landau theory of classical phase transition fails in order to
describe these phases. Ground state of such phases is a highly
entangled state and the excitations above the ground state have a
topological nature which mirrored in their exotic statistics.

Among the models with topological properties, the Kitaev or toric
code\cite{toric} has been extensively studied. Ground state
degeneracy depends on the genus or handles of the manifold where the
model is defined on, and there is a gap which separated the ground
state subspace from the excited states. The ground state degeneracy
can not be lifted by any local perturbations which underlines the
Kitaev's model as a testground for fault-tolerant quantum
computations \cite{toric}. The ground state of the Kitaev's model is
indeed stabilized by group generated by a set of local operators
called \emph{plaquette} and \emph{star } operators making it useful
as a quantum error correcting code\cite{gottesman}. The information
is encoded in the ground state subspace which is topologically
protected making it robust as a quantum memory \cite{dennis}. For
this model any bipartition of the lattice has non-zero entanglement
which manifests the ground state is generically multipartite
entangled\cite{hamma}. Bipartite entanglement scales with the
boundary of the subsystem showing that the entanglement between two
parts of the system depend only on the degrees of freedom living on
the boundary which is a manifestation of the holographic character
of the entanglement
entropy\cite{takayanagi}.\\

For topological models a satisfactory connection between topological
order and entanglement content of a model has been established via
introducing the concept of the \emph{topological entanglement
entropy (TEE)}\cite{kitaev,levin} which is a universal
quantity with topological nature.\\

Another resource with topological protection character is called
\emph{color code}.  An interplay between color and homology provides
some essential features, for example a particular class of
two-dimensional color codes with colored borders will suppress the
need for selective addressing to qubits through  implementation of
Clifford group\cite{bombin1}.The number of logical qubits which are
encoded by a two dimensional color code are twice
than the toric code \cite{bombin2}defined on the compact surface.\\
In this paper we study the entanglement properties of a
 color code defined on a two-dimensional lattice, the so called 2-colex \cite{bombin3}. We consider
different bipartitions of the lattice and evaluate the entanglement
entropy between them. The connection between the topological nature
of the code and the entanglement
entropy is also discussed.\\
The structure of the paper is as follows. In sec(II) the basic
notions of a color code on the surface has been introduced. In
sec(III) based on the stabilizer structure of the protected code
space the reduced density matrix which is needed for evaluating the
entanglement entropy has been calculated. Then in sec(IV) the
entanglement entropy for different bipartitions is calculated. In
sec(V) another class of color code on plane and its entanglement
properties are introduced, and the final section (VI) has been
devoted to the conclusions.

%%%%%%%%%%%%%%%%%%%%%%%%%%%%%%%%%%%%%%%%%%%%%%%%%%%%%%%%%%%%%%%%%
\section{preliminaries on topological color code}
\begin{figure}
\begin{center}
\includegraphics[width=8cm]{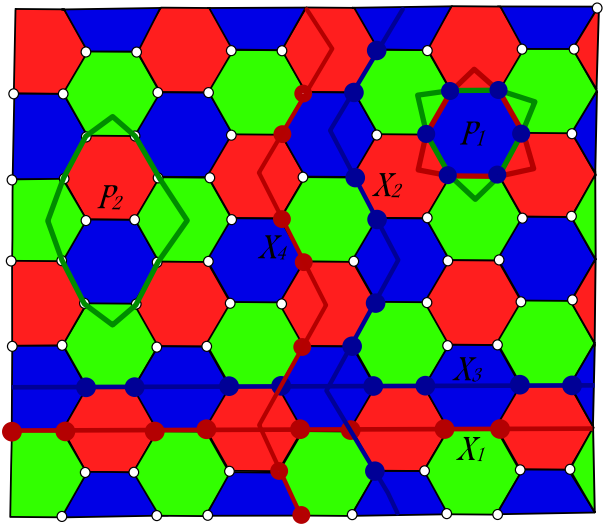}
\caption{(color online) A piece of 2-colex which has been defined on
the hexagonal lattice with periodic boundary conditions. Lightest to
darkest hexagons: green, red and blue, respectively. All circles
stand for qubits living at the vertices. A hexagon $(P_{1})$, say
blue, can be denoted by either red or green closed boundary string
and the product of two neighboring plaquettes, say red and blue,
corresponds to a green string $(P_{2})$. Non-contractible loops
($X_{1}$,$X_{2}$,$X_{3}$,$X_{4}$) for topological color codes which
determine the structure of the encoded subspace $\mathcal{C}$. A
colored string connect plaquettes with the same color.} \label{fig1}
\end{center}
\end{figure}

In this paper we consider a class of topological quantum
error-correction code defined on the lattice, the so called
topological color codes (TCC)\cite{bombin1}. The local degrees of
freedom are spin-$1/2$ with the bases of the Hilbert space
\emph{$C^{2}$}. The lattice we consider composed of vertices, links
and plaquettes. Each vertex stands for a local spin. Three links
meet each other at a vertex and no two plaquettes with same color
share the same link. We suppose this color structure is denoted by
the notation TCC\{\emph{V},\emph{E},\emph{P}\} where the
\emph{V},\emph{E},\emph{P} denote the set of vertices, edges and
plaquettes, respectively. For simplicity we define the model on the
regular hexagonal lattice on the torus, i.e imposing periodic
boundary conditions as shown in Fig.(\ref{fig1}). There is a
subspace $\mathcal{C}\subset \mathcal{H}$ which is topologically
protected. The full structure of this subspace and its properties
are determined by definition of stabilizer group. The stabilizer
group is generated by a set of plaquette operators. For each
plaquette we attach the following operators which are product of a
set of Pauli operators of vertices around a plaquette: \bea
\label{eq1} \Omega^{C}_{p}=\bigotimes_{v\in p}\Omega^{C}_{v}~;~~
\Omega=\emph{X}, \emph{Z}~~ ,
~~C=\mathrm{Red},\mathrm{Green},\mathrm{Blue}.\eea

For a generic plaquette, say blue plaquette $P_{1}$ in
Fig.(\ref{fig1}) we can identify green and red strings which are the
boundary of the plaquette. It is natural to think of product of
different plaquette operators which may produce a collection of
boundary operators. For example as is shown in Fig.(\ref{fig1}) the
product of two neighboring plaquettes, say red and blue ones,
correspond to a green string $P_{2}$ which is a boundary of two
plaquettes. All string operators produced in this way are closed.
Since all closed strings share either nothing or even number of
vertices, they commute with each other and with plaquettes . In
addition to closed boundary operators, there are other closed string
which are no longer the product of the plaquette operators. These
closed string form the fundamental cycles of the manifold in which
the lattice is defined on and have a character of color. Number of
these closed loops depends on the genus of the manifold where the
lattice defined on. For the torus with $\mathrm{g}=1$ there are two
such cycles which are non-contractible loops in contrary to the
closed boundary strings which are homotopic to the boundary of a
plaquette. For the topological color codes these non-contractible
loops are shown in Fig.(\ref{fig1}). For every homology class of the
torus there are two closed strings each of one color, say red and
blue. Red string connects red plaquettes and so on. Note that a
generic string, say green, can be produced by product of the red and
blue strings when they are suitably chosen. In fact since every
Pauli matrix squares identity, at a vertex (qubit) which two
homologous strings cross each other they cancel each other. Indeed
there is an interplay between color and homology class of the model.
One can define a nontrivial closed string as follows:
 \bea \label{eq2} \mathcal{S}^{C\Omega}_{\mu}=\bigotimes_{i\in I}\Omega_{i},\eea
where \emph{I} indexed the set of spins on a generic string, $\mu$
stands for the homology class of the torus and $\Omega$ is the $X$
or $Z$ Pauli spin operators. Closed non-contractible loops turn on
to form bases for the encoded logical operators of topological code.
To more clarify this, we label different loops as:

\bea \no X_{1}\longleftrightarrow
\mathcal{S}^{RX}_{2}~,X_{2}\longleftrightarrow
\mathcal{S}^{BX}_{1}~,X_{3}\longleftrightarrow
\mathcal{S}^{BX}_{2}~,X_{4}\longleftrightarrow
\mathcal{S}^{RX}_{1},\eea

\bea \label{eq3} Z_{1}\longleftrightarrow
\mathcal{S}^{BZ}_{1}~,Z_{2}\longleftrightarrow
\mathcal{S}^{RZ}_{2}~,Z_{3}\longleftrightarrow
\mathcal{S}^{RZ}_{1}~,Z_{4}\longleftrightarrow
\mathcal{S}^{BZ}_{2}~.\eea

These operators form a 4-qubit algebra in $\mathcal{H}_{2}^{4}$, so
it manifests a 16-dimensional subspace for the coding space
$\mathcal{C}$ which is topologically protected. On the other hand,
the topological color code on the torus with $\mathrm{g}=1$ encodes
four qubits. Now we move to construct the explicit
form of the states of the subspace $\mathcal{C}$.\\
The above construction for the string operators in Eq.(\ref{eq3})
can be extended to an arbitrary manifold with genus $\mathrm{g}$.
For such manifold the coding space spanned with $2^{4\mathrm{g}}$
vectors. Note that for toric code (white and dark
code)\cite{bombin2} when is embedded in the same manifold with genus
$\mathrm{g}$, the coding space will span with $2^{2\mathrm{g}}$
vectors\cite{hamma} which explicitly shows that the color codes have
richer structure than the toric codes\cite{bombin2}.

%%%%%%%%%%%%%%%%%%%%%  Fig.1   %%%%%%%%%%%%%%%%%%%%%%%%%%%%%%%%%%%%%%%%%%%%%%%%%%%%%%%%%%%
%%%%%%%%%%%%%%%%%%%%%%%%%%%%%%%%%%%%%%%%%%%%%%%%%%%%%%%%%%%%%%%%%%%%%%%%%%%%%%%%%%%%%%%%%%%

\subsection{Stabilizer Formalism}
 The protected subspace $\mathcal{C}$ is spanned by the state
vectors which are stabilized by all element of stabilizer group,
i.e. a subset of Pauli group. Let $\mathcal{U}$ be a set of
generators of the stabilizer group and its elements are denoted by
$\mathcal{M}$. So this subspace is

\bea \label{eq4} \mathcal{C}=\{|\psi\rangle~:~~~~
\mathcal{M}|\psi\rangle=|\psi\rangle ~~~ \forall \mathcal{M}\in
\mathcal{U}\}~~. \eea

Let $G$ be the group constructed by the generators of spin-flip
plaquette operators, i.e $X^{C}_{p}$. The cardinality of the group
is then $|G|=2^{|P|-2}$ where $|P|$ stands for the total number of
plaquettes. Note that all plaquettes are not independent since the
product of all plaquettes with the same color represents the same
action in the group, namely
$\prod_{R}X^{C}_{p}=\prod_{B}X^{C}_{p}=\prod_{G}X^{C}_{p}=X^{\bigotimes
|V|}$ where $|V|$ stands for the number of all vertices. By starting
from an initial vacuum state, say $|0\rangle^{\bigotimes |V|}$ where
$Z|0\rangle=|0\rangle$, one can construct a state vector in the
Hilbert space which is stabilized by the group elements. This state
vector is a superposition of all elements of the stabilizer group
with equal weights. From the previous arguments, it is convenient to
denote it by $|0000\rangle$ which has the following form \bea
\label{eq5} |0000\rangle=|G|^{-1/2}\sum_{g\in G}
g|0\rangle^{\bigotimes |V|}, \eea where the $g$ is an element of the
stabilizing group, i.e $g=\bigotimes_{p\in P}X^{r_{p}}_{p}$ where
$r_{p}=0(1)$ corresponds to the plaquette operator $X_{p}$ appearing
(not appearing) in the element group $g$. There are many elements of
the Pauli group that commute with the all elements of the stabilizer
group but are not actually in $G$, and it is defined as the
centralizer of $G$ in Pauli group. Since elements of Pauli group
either commute or anticommute, the centralizer is actually equal to
the normalizer of $G$ in Pauli group. Considering the
non-contractible loops, the normalizer of stabilizer group can be
obtained by product of stabilizer group and group of
non-contractible loop operators. Let denote it by
$\bar{G}=\mathcal{A}\cdot G$ where the group $\mathcal{A}$ generated
by non-contractible loop operators, i.e. \bea \label{eq6}
\mathcal{A}=\{\prod^{\mu=2\mathrm{g}}_{\mu=1,C=B,R}(S^{CX}_{\mu})^{r_{\mu
c}}~~,~~r_{\mu c}=0,1\}~~.\eea

So, the group $\bar{G}$ reads \bea \label{eq7}
\bar{G}=\{\prod^{\mu=2\mathrm{g}}_{\mu=1,C=B,R}(S^{CX}_{\mu})^{r_{\mu
c}} \cdot G~~,~~r_{\mu c}=0,1\}~~. \eea

Note that $G\subseteq \bar{G}$. The group $\bar{G}$ is the
normalizer of $G$ in Pauli group, so the normal subgroup $G$ divides
the group $\bar{G}$ into $2^{4\mathrm{g}}$ cosets. The cardinality
of group will be $|\bar{G}|=2^{|P|+4\mathrm{g}-2}$. Therefore the
protected subspace $\mathcal{C}$ is spanned by $2^{4\mathrm{g}}$
states which correspond to different cosets. In fact, elements in
$\bar{G}-G$ take one encoded state of the stabilized subspace to
another encoded state without leaving the stabilized subspace. For
the states of the stabilized subspace we have, by construction:

\bea \label{eq8} \mathcal{C}=\{~|ijkl\rangle:
~~~|ijkl\rangle=X^{i}_{1}X^{j}_{2}X^{k}_{3}X^{l}_{4}|0000\rangle~\},
\eea where $X_{1},X_{2},X_{3},X_{4}$ are defined in Eq.(\ref{eq3})
and $i,j,k,l=0,1$. These topological non-trivial string operators
can take one state of the coding space to another one, and any error
of this type will not be detectable. A generic state can be a
superposition of different vectors of coding space as follows:

\bea \label{eq9} |\Psi\rangle=\sum_{i,j,k,l}a_{i,j,k,l}|ijkl\rangle
~~~~,~~~~\sum_{i,j,k,l}|a_{i,j,k,l}|^{2}=1~~.\eea

%%%%%%%%%%%%%%%%%%%%%%%%%%%%%%%%%%%%%%%%%%%%%%%%%%%%%%%%%%%%%%%%%%%%%%%%%%%%%%%

\subsection{Protected Subspace as a Ground State Subspace}

From the practical point of view it is important to find a state of
quantum many-body system for implementation of universal quantum
computation. We can provide a construction in which the protected
subspace be the ground sate of a local
Hamiltonian\cite{raussendorf}. The subspace $\mathcal{C}$ is the
ground state of a following exactly solvable Hamiltonian , i.e.

\bea \label{eq10} H=-\sum_{p \in P}X_{p}-\sum_{p \in P}Z_{p}~~.\eea

The ground state of this Hamiltonian is $2^{4\mathrm{g}}$-fold
degenerate and topologically protected from local errors. Different
states of the ground state subspace as it is clear from
Eq.(\ref{eq8}) are obtained by product of non-contractible colored
strings. Each state is characterized by a set of topological numbers
which are sum of all $z_{v}$ modula $2$ along non-contractible
loops. For example for a generic state $|ijkl\rangle$ in
Eq.(\ref{eq8}) the topological numbers are $\sum_{v\in
I_{1}^{B}}z_{v}=i$, $\sum_{v\in I_{2}^{R}}z_{v}=j$, $\sum_{v\in
I_{1}^{R}}z_{v}=k$ and $\sum_{v\in I_{2}^{B}}z_{v}=l$, where the
summations are evaluated in mod $2$, $z_{v}$ stands for the
$z$-component of $Z$-Pauli matrix of vertex $v$ and $I_{\mu}^{C}$
identifies a set of vertices (qubits) on a closed non-contractible
loop with homology $\mu$ and color $C$. An excited state will arise
when any of the stabilizing conditions in Eq.(\ref{eq4}) is
violated. In the energy unit which is defined by Hamiltonian
Eq.(\ref{eq10}),
 the first excited state will be separated from the
ground state by a gap of value $2$.

%%%%%%%%%%%%%%%%%%%%%%%%%%%%%%%%%%%%%%%%%%%%%%%%%%%%%%%%%%%%%%%%%%%%%%%%%%%%%%
\section{reduced density matrix and von Neumann entropy}

In this section we turn on to calculate the entanglement properties
of the topological color codes. We consider a generic bipartition of
the system into subsystems $A$ and $B$. Let $\Sigma_{A}$ and
$\Sigma_{B}$ to be the number of plaquette operators acting solely
on $A$ and $B$, respectively, and let $\Sigma_{AB}$ to stand for the
number of plaquette operators acting simultaneously on $A$ and $B$,
i.e these are boundary operators. We focus on the entanglement
entropy between two partitions $A$ and $B$ of the system. To this
end, first the reduced density operator of the one subsystem is
evaluated and then the entanglement entropy is measured. If the
state of the system is in an equal superposition of the elements of
group \emph{G}, the reduced density matrix for a subsystem, say
\emph{A}, has the following form \cite{hamma}

\bea \label{eq11} \rho_{A}=\frac{d_{B}}{|G|}\sum_{g\in
G/G_{B},\tilde{\rho}\in
G_{A}}g_{A}|0_{A}\rangle\langle_{A}0|g_{A}\tilde{g}_{A}\eea where
$G_{A}$ and $G_{B}$ are subgroups of $G$ which act trivially on
subsystems \emph{B} and \emph{A}, respectively and $d_{A}$ and
$d_{B}$ are their cardinality. So the von Neumann entropy is

\bea \label{eq12} S_{A}=\log_{2}|G_{AB}|\eea where
$G_{AB}=\frac{G}{G_{A}G_{B}}$. It is a simple task to show that all
states of the coding space have the same entanglement entropy. To
show this, let $X(t)=X^{i}_{1}X^{j}_{2}X^{k}_{3}X^{l}_{4}$ in which
$t=(i,j,k,l)$ is a binary vector. So, for a generic state in the
coding space we have: $|t\rangle=X(t)|0\rangle$. Moreover the string
operators $X(t)$ can be decomposed as $X(t)=X(t)_{A} \bigotimes
X(t)_{B}$. Therefore we obtain
\begin{widetext}

\bea \label{eq12} \rho_{A}(|t\rangle)=Tr_{B}(|t\rangle\langle
t|)=Tr_{B}(X(t)|0\rangle\langle0|X(t))=X(t)_{A}Tr_{B}(|0\rangle\langle0|)X(t)_{A}=X(t)_{A}\rho_{A}(|0\rangle)X(t)_{A}~.\eea

\end{widetext}
This implies that, by using the ancilla $S_{A}(t)=\lim_{n
\rightarrow 1}
\partial_{n} Tr[\rho^{n}_{A}]$, the entanglement entropy for a state $|t\rangle$ is $S_{A}(t)=S_{A}(0)$.

%%%%%%%%%%%%%%%%%%%%%%%%%%%%%%%%%%%%%%%%%%%%%%%%%%%%%%%%%%%%%%%%%%%%%%%%%%%%%%%%%%%%%%%%%%%%%%%%%%%%%%%%%%%%%%%%%%%%%%%%%%%%%%%%%%%%

\section{entanglement of TCC for various bipartitions}
In this section we design different spin configurations as
subsystems and then evaluate their entanglement with its
complementary.

\subsection{One Spin}
As first example we consider the entanglement between one spin and
remaining ones of the lattice. In this case there is no closed
boundary operators acting exclusively on the spin, i.e subsystem
\emph{A}. So the reduced density matrix $\rho_{A}$ is diagonal as
follows:

\bea \label{eq13} \rho_{A}=f^{-1}\sum_{g\in
G/G_{B}}g_{A}|0_{A}\rangle\langle_{A}0|g_{A}\eea

where $f=|G/G_{B}|$ is the number of operators which act freely on
the subsystem \emph{A}. As it is clear there are three plaquette
operators which act freely on one spin since every spin in the color
code is shared by the three plaquettes (See Fig.(\ref{fig1})). This
leads to $f=2^{3}=8$. Only half of operators of quotient group
leading to flip the spin, i.e. they will have trivial effect unless
three plaquettes act either individually or altogether. Therefore
the reduced density matrix is

%=\frac{1}{8}(4\sigma^{x}_{A}|0_{A}\rangle\langle_{A}0|\sigma^{x}_{A}+4|0_{A}\rangle\langle_{A}0|)
\bea \label{eq14} \rho_{A}=
\frac{1}{2}(|1_{A}\rangle\langle_{A}1|+|0_{A}\rangle\langle_{A}0|).\eea

Thus, in topological color code every spin is maximally entangled
with other spins. For the surface code (Kitaev model) a single spin
is also maximally entangled with other spins \cite{hamma}.

\subsection{Two Spins}

We calculate the entanglement between two spins. For this case also
there is not any closed boundary operator with nontrivial effect on
two spins, so the reduced density matrix reads

\bea
\label{eq15}\rho_{A}=\frac{1}{4}(|11\rangle\langle11|+|10\rangle\langle10|+|01\rangle\langle01|+|00\rangle\langle00|).\eea

The entanglement between two spins which is measured by the
Concurrence\cite{wootters} vanishes. While each spin is maximally
entangled with others, two spins are not entangled which is a
manifestation of the fact that the topological color code is
genuinely multipartite entangled like the surface codes.

\subsection{Colored Spin Chain}
In this case we aim to know how much a closed colored spin chain
winding the torus nontrivially, say blue or red in Fig.(\ref{fig1})
is entangled with rest of the lattice. For the sake of clarity and
without loss of generality we consider a hexagonal lattice with
$|P|=3k\times4k$ plaquettes, where $k$ is an integer number as
$1,2,3...$, for example in Fig.(\ref{fig1}) $k=2$. This choice makes
the color code more symmetric. So, the number of plaquettes with a
specific color, say red, is $(2k)^{2}$, and the number of spins that
the colored chain contains is $4k$. Let the system be in the
$|0000\rangle$. For this state there is not any closed boundary
which exclusively acts on the chain. Therefor the reduce density
matrix is diagonal and the number of plaquette operators which act
independently on the subsystem \emph{B} is \bea
\label{eq16}\Sigma_{B}=(|P|-2)-(3\times2k)+(2k+1)\eea where the
third term is the number of constrains on the plaquette operators
acting on spin chain. In fact these are collective operators, i.e
product of boundary plaquette operators between \emph{A} and
\emph{B}which act solely on \emph{B}. With these remarks in hand,
the entanglement entropy becomes

 \bea
\label{eq17}S_{A}=\log\frac{|G|}{d_{B}}=4k-1~.\eea

It is instructive to compare the obtained entanglement entropy in
Eq.(\ref{eq17}) with the entanglement of the spin chain in the
Kitaev's model. For the latter case when the model defined on the
square lattice on a compact surface, i.e torus, there exists a
rather similar scaling with the number of qubits living in the
chain\cite{hamma}.

In both cases the entanglement entropy in Eq.(\ref{eq17}) can be
understood from the fact that only configurations with even number
of spin flips of the chain allow in the ground state structure.
Plaquettes which are free to act on the spin chain only able to flip
even number of spins of the chain. The same arguments can be applied
for other ground states $|ijkl\rangle$ where for all of them the
entanglement entropy is given by Eq.(\ref{eq17}). By inspection of
other ground states we see that there are $2^{3}=8$ state in which
even number of spins in the spin chain have been flipped. These
states are

\bea \no
|0000\rangle~,X_{1}|0000\rangle~,X_{3}|0000\rangle~,X_{4}|0000\rangle~,X_{1}X_{3}|0000\rangle,\eea

\bea \label{eq18}
X_{1}X_{4}|0000\rangle~,~X_{3}X_{4}|0000\rangle~,~X_{1}X_{3}X_{4}|0000\rangle.\eea

Besides, there are $2^{3}=8$ states with an odd number of spin
flipped for the spin chain which have been listed below:

\bea \no
X_{2}|0000\rangle~~~,~~~X_{2}X_{1}|0000\rangle~~~,~~~X_{2}X_{3}|0000\rangle,\eea

\bea \no
X_{2}X_{4}|0000\rangle~,X_{2}X_{1}X_{3}|0000\rangle~,X_{2}X_{1}X_{4}|0000\rangle,\eea

\bea \label{eq19}
X_{2}X_{3}X_{4}|0000\rangle~,X_{2}X_{1}X_{3}X_{4}|0000\rangle.\eea

Now we consider a generic state and calculate the entanglement
entropy between the colored spin chain and rest of the lattice. The
density matrix $\rho=|\Psi\rangle\langle\Psi|$ is \bea
\label{eq20}\rho=\sum_{ijkl,mnpq}a_{ijkl}\bar{a}_{mnpq}X^{i}_{1}X^{j}_{2}X^{k}_{3}X^{l}_{4}\rho_{0}X^{m}_{1}X^{n}_{2}X^{p}_{3}X^{q}_{4}.\eea

With the explanations we made above the eigenvalues of the reduced
density matrix $\rho_{A}$ for the even number of spin flipped is as
follows:
 \bea \no \frac{d_{B}}{|G|}(|a_{0000}|^{2}+|a_{1000}|^{2}+|a_{0010}|^{2}+|a_{0001}|^{2}+|a_{1010}|^{2}\eea

\bea \label{eq21}
+|a_{1001}|^{2}+|a_{0011}|^{2}+|a_{1011}|^{2})=\frac{d_{B}}{|G|}\alpha
\eea

where $\frac{|G|}{d_{B}}=2^{4k-1}$. Note that the number of these
eigenvalues is $2^{4k-1}$. Through the similar arguments we see that
the eigenvalues of reduced density matrix of the spin chain with odd
number of spin flipped  is $\frac{d_{B}}{|G|}(1-\alpha)$ and the
number of these eigenvalues is $2^{4k-1}$. So, we can calculate the
entanglement entropy as follows:

\begin{widetext}
\bea
\label{eq22}S_{A}=-\sum^{2^{4k}}_{i=1}\lambda_{i}\log\lambda_{i}=-2^{4k-1}\frac{d_{B}}{|G|}\alpha\log\frac{d_{B}}{|G|}\alpha-2^{4k-1}\frac{d_{B}}{|G|}(1-\alpha)\log(\frac{d_{B}}{|G|}(1-\alpha))=4k-1+H(\alpha)\eea
\end{widetext}

where $H(x)=-x\log x-(1-x)\log(1-x)$.\\ What about the entanglement
of an open string? the open strings in the color code stand for the
errors and map the coding space into an orthogonal one. These open
strings anticommute with plaquette operators which share in odd
number of vertices. For an open string which contains $k'$ qubits
the entanglement entropy is: $S_{A}=\log\frac{|G|}{d_{B}d_{A}}=k'$,
i.e an open string will be maximally entangled with rest of the
system.

\subsection{Red Strings Crossing}

As an another bipartition we select all spins on the two red strings
with different homologies, say $X_{1}$ and $X_{4}$ in
Fig.(\ref{fig1}), which have $2\times4k=8k$ spins. Let system be in
the $|0000\rangle$ state. There is not any closed boundary operator
acting solely on \emph{A}. Considering the total number of
independent plaquette operators acting solely on subsystem \emph{B},
the entanglement entropy then reads

\bea \label{eq22}S_{A}=8k-1 .\eea

Again we see that there are only  configurations with an even number
of spin flipped in the construction of the state $|0000\rangle$. For
a generic state as in Eq.(\ref{eq20}) where we can realized all
states with either even or odd number of spin flips, the
entanglement entropy will be

\bea \label{eq22}S_{A}=8k-1+H(\alpha)~.\eea

\subsection{Red and Blue strings crossing}
By this partition we mean the spins on the  $X_{1}$ and $X_{2}$
strings in Fig.(\ref{fig1}). The subsystem \emph{A} contains $8k-1$
spins and again there is no closed boundary acting on it.
Enumerating the independent plaquette operators acting on the
subsystem \emph{B}, the entanglement entropy is

\bea \label{eq23}S_{A}=8k-3~.\eea

\subsection{Two Parallel Spin Chains}
Two parallel spin chains, say red and blue ($X_{1}$ and $X_{3}$) in
Fig.(\ref{fig1}) or equivalently a green string, contain
$2\times4k=8k$ spins and again there is not any closed string acting
on it nontrivially. $8k$ plaquettes act freely on subsystem $A$, but
there are some constraints on them which arise from the product of
plaquettes in which leave the subsystem $A$ invariant and makes
collective operator which acts solely on \emph{B}. The product of
blue and green plaquettes of those plaquettes which are free to act
on $A$ will produce a red string leaving $A$ invariant. The same
product hold for the red and green plaquettes. So the entanglement
entropy reads

 \bea \label{eq24}S_{A}=8k-2~.\eea
Indeed this is entanglement entropy for a green string. However, the
number of spins it contains is twice than the strings we discussed
in Subsec(C).

%%%%%%%%%%%%%%%%%%%%%  Fig.2   %%%%%%%%%%%%%%%%%%%%%%%%%%%%%%%%%%%%%%%%%%%%%%%%%%%%%%%%%%%
\begin{figure}
\begin{center}
\includegraphics[width=7cm]{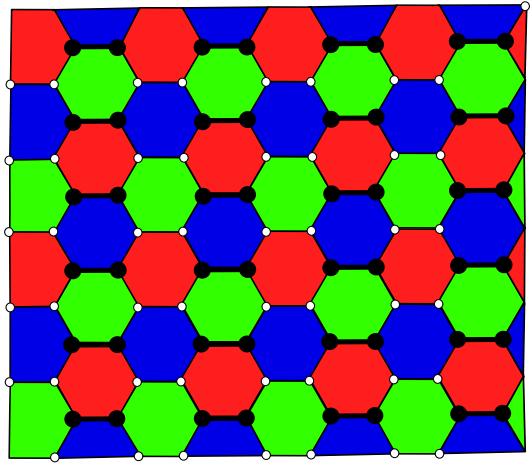}
\caption{(color online) A set of vertical spin ladders (black solid
circles) on 2-colex as a subsystem for calculating entanglement
entropy.} \label{fig2}
\end{center}
\end{figure}
%%%%%%%%%%%%%%%%%%%%%%%%%%%%%%%%%%%%%%%%%%%%%%%%%%%%%%%%%%%%%%%%%%%%%%%%%%%%%%%%%%%%%%%%%%%

\subsection{Spin Ladder}
A set of vertical spin ladders has been shown in Fig.(\ref{fig2}).
In this subsection only one of them has been considered as subsystem
$A$. Let the system be in the state $|0000\rangle$. In this case the
subsystem \emph{A} contains $2\times3k$ spins, and the total number
on independent plaquette acting on subsystem \emph{B} is
$(|P|-2)-3\times3k+3k$. Since there are not any closed string acting
solely on \emph{A}, the entanglement entropy reads

\bea \label{eq25}S_{A}=6k~.\eea

So, it is clear that this spin ladder has maximum entanglement with
the rest of the system and its state has the maximum mixing, namely

\bea \label{eq26}\rho_{A}=2^{-6k}\mathbf{I}_{6k\times 6k}~.\eea

\subsection{Vertical Spin Ladders}

Now we consider all vertical spin ladders has been depicted in
Fig.(\ref{fig2}) and again we suppose the system is in the state
$|0000\rangle$. There is not any plaquette operator which acts
solely on subsystem \emph{A}. However, there are several specific
product of plaquette operators producing closed strings in which act
solely on \emph{A}. For example the product of plaquettes which are
in a vertical column will produce a closed string acting on one of
the subsystems. Both subsystems \emph{A} and \emph{B} are symmetric
with respect to each other. So, the number of closed strings which
act only on \emph{A} and \emph{B} will be $d_{A}=d_{B}=2^{2k}$.
Finally the entanglement entropy reads

\bea \label{eq27}S_{A}=12k^{2}-4k-2~.\eea

%%%%%%%%%%%%%%%%%%%%%  Fig.3  %%%%%%%%%%%%%%%%%%%%%%%%%%%%%%%%%%%%%%%%%%%%%%%%%%%%%%%%%%%
\begin{figure}
\begin{center}
\includegraphics[width=7cm]{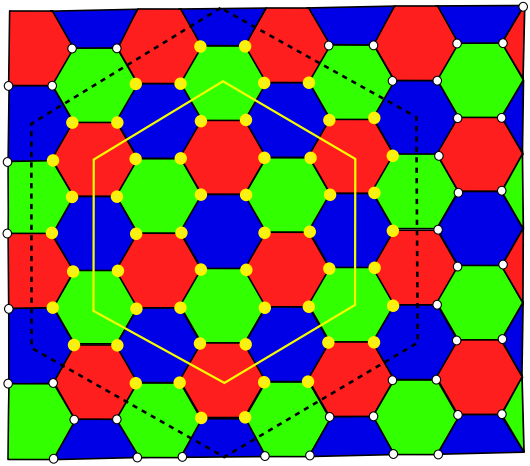}
\caption{(color online) A manifestation of a hexagonal disk as a
subsystem. A set of plaquettes is chosen in such a way that form a
large hexagon (the yellow line), and all large yellow (light) spins
are in the subsystem . The dashed black line depicts the set of
plaquettes acting simultaneously on both subsystems.} \label{fig3}
\end{center}
\end{figure}
%%%%%%%%%%%%%%%%%%%%%%%%%%%%%%%%%%%%%%%%%%%%%%%%%%%%%%%%%%%%%%%%%%%%%%%%%%%%%%%%%%%%%%%%%%%

\subsection{A Hexagonal Disk}

In this case we adapt a situation in which a set of plaquettes
intuitively forms a hexagon as shown in Fig.(\ref{fig3}). In this
figure the dashed hexagon demonstrates a set of plaquettes lying
between two subsystems. We suppose that the number of plaquettes
crossed by the one edge of the hexagon be $n$. Now we can enumerate
the number of plaquettes which act on $A$ in terms of $n$, i.e.
$\Sigma_{A}=3n(n-1)+1$. The total number of spins of the subsystem
\emph{A} is $6n^{2}$. There are $\Sigma_{AB}=6n$ plaquettes which
act between \emph{A} and \emph{B}. Let the system be in the state
$|0000\rangle$. With these realizations of the subsystems and since
$|P|=\Sigma_{A}+\Sigma_{B}+\Sigma_{AB}$ we can turn on to calculate
the entanglement entropy between two subsystems as follows:

\bea \label{eq28}S_{A}=\log_{2}
2^{\Sigma_{AB}-2}=\Sigma_{AB}-2=6n-2~.\eea

We can relate this entropy to the perimeter of the subsystem
\emph{A} by choosing the hight of a plaquette as unit. The perimeter
of \emph{A} is $\partial A=6n$ and the entropy then will be

\bea \label{eq29}S_{A}=\partial A-2~. \eea Generally for a irregular
lattice the entropy for a convex shape will be $S_{A}=\kappa\partial
A-\gamma$ where the coefficient $\kappa$ is a nonuniversal constant
depending on the shape of the region while the constant $\gamma=2$
will be universal and has a topological nature. So, we see that the
entanglement entropy for the topological color code scales with the
boundary of the subsystem which is a manifestation of so
called \emph{area law}\cite{takayanagi,plenio,bousso}.\\
The latter relation for we obtained for the entanglement entropy is
consistent with derivation of A. Kitaev \emph{et al.}\cite{kitaev}
and M. Levin \emph{et al.}\cite{levin}  where they proposed for a
massive topological phase there is a topological subleading term for
the entanglement entropy which is related to the quantum dimension
of the abelian quasiparticles. The total quantum dimension for the
topological color code will be

\bea \label{eq30}\gamma=\log\mathcal{D} ~~,~~~\mathcal{D}=4~.\eea

The quantity $\mathcal{D}^{2}$ is the number of topological
superselection sectors of abelian anyons. For Kitaev's toric code
there are only four such sectors leading to $\gamma=\log2$. So the
total quantum dimension in the topological color code is bigger than
the toric code. The abelian phase of the toric code is characterized
via appearing a global phase for the wavefunction of the system by
winding an electric (magnetic) excitation around a magnetic
(electric) excitation \cite{kitaev}. In the case of color code the
excitations are colored and they appear as end points of open
colored strings of a shrunk lattice\cite{bombin3}. Winding a colored
$X$-type excitation around the $Z$-type one with different color
gives an overall factor (-) for the wavefunction of the model, i.e
the phase is abelian. Therefore the contribution of the color to the
excitations makes the abelian phase of the color code richer than
the toric code with a bigger quantum dimension.

Now let the system be in a generic state such as Eq.(\ref{eq20}).
The reduced density matrix then reads
\begin{widetext}
\bea \label{eq31} \rho_{A}=\sum_{ijkl,mnpq}a_{ijkl}\bar{a}_{mnpq}
Tr_{B}(X^{i}_{1}X^{j}_{2}X^{k}_{3}X^{l}_{4}\rho_{0}X^{m}_{1}X^{n}_{2}X^{p}_{3}X^{q}_{4})~.\eea
\end{widetext}
Some remarks are in order. For two different nontrivial closed
string operators but with the same homology and color , say $X$ and
$X'$, they will have same support and are related via a trivial
closed string $g\in G$, i.e $X'=gX$. Since $[g,X]=0$, two string
will have the same effect on the $\rho_{0}$, i.e.
$X'\rho_{0}=gX\rho_{0}=X\rho_{0}$. We can exploit this property of
string operators in order to choose the strings appearing in
Eq.(\ref{eq31}) in which they do not cross subsystem $A$. Indeed the
strings $X_{i}$ act only on the subsystem $B$, i.e
$X_{iA}=\mathbf{I}_{A}$. This leads to
$Tr_{B}(X_{i}\rho_{0}X_{i})=Tr_{B}(\rho_{0})$ for $i=1,2,3,4$ and
$Tr_{B}(X^{i}_{1}X^{j}_{2}X^{k}_{3}X^{l}_{4}\rho_{0}X^{i}_{1}X^{j}_{2}X^{k}_{3}X^{l}_{4})=Tr_{B}(\rho_{0})$
for $i,j,k,l=0,1$. Other cases will be zero. To clarify this, let
for example consider the simple case $Tr_{B}(X_{1}\rho_{0})$ where
we have

\begin{widetext}
 \bea \label{eq32} Tr_{B}(X_{1}\rho_{0})=\sum_{g,g'\in
G}g_{A}|0_{A}\rangle\langle_{A}0|g'_{A}\sum_{g''_{B}}\langle_{B}0|g''_{B}X_{4}g_{B}|0_{B}\rangle\langle_{B}0|g'_{B}g''_{B}|0_{B}\rangle
=\sum_{g,g'\in
G}g_{A}|0_{A}\rangle\langle_{A}0|g'_{A}\langle_{B}0|g'_{B}X_{B}g_{B}|0_{B}\rangle
.\eea \end{widetext}

On the other hand from the group property $g'=g\tilde{g}$ the above
expression becomes

\bea \label{eq33} Tr_{B}(X_{1}\rho_{0})=\sum_{g,\tilde{g}\in
G}g_{A}|0_{A}\rangle\langle_{A}0|g_{A}\tilde{g}_{A}\langle_{B}0|\tilde{g}_{B}X_{B}|0_{B}\rangle\eea

which implies that $\tilde{g}_{B}=X_{B}$. Otherwise the above
expression will become zero. The fact
$\tilde{g}=\tilde{g}_{A}\otimes\tilde{g}_{B} \in G$ explicitly
implies that the operator $\tilde{g}_{A}$ must be a non-contractile
strings which is impossible since it must act solely on $A$. So we
will leave with $Tr_{B}(X_{1}\rho_{0})=0$. The same arguments may
also apply for other cases. Finally the entanglement for a generic
state of the system and a convex region will become

\bea \label{eq34} \rho_{A}=Tr_{B}(\rho_{0})~~~,~~~~S_{A}=\partial
A-2~~.\eea

%%%%%%%%%%%%%%%%%%%%%  Fig.4  %%%%%%%%%%%%%%%%%%%%%%%%%%%%%%%%%%%%%%%%%%%%%%%%%%%%%%%%%%%
\begin{figure}
\begin{center}
\includegraphics[width=7cm]{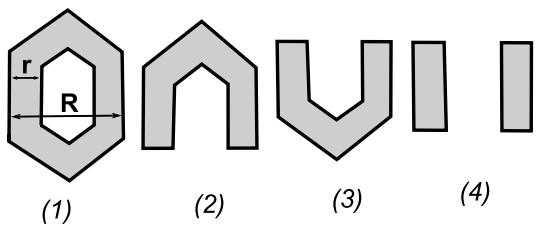}
\caption{(color online) Different bipartitions of the lattice in
order to drop the bulk and boundary effect through the definition of
topological entanglement entropy.} \label{fig4}
\end{center}
\end{figure}
%%%%%%%%%%%%%%%%%%%%%%%%%%%%%%%%%%%%%%%%%%%%%%%%%%%%%%%%%%%%%%%%%%%%%%%%%%%%%%%%%%%%%%%%%%%

%%%%%%%%%%%%%%%%%%%%%  Fig.5  %%%%%%%%%%%%%%%%%%%%%%%%%%%%%%%%%%%%%%%%%%%%%%%%%%%%%%%%%%%
\begin{figure}
\begin{center}
\includegraphics[width=8cm]{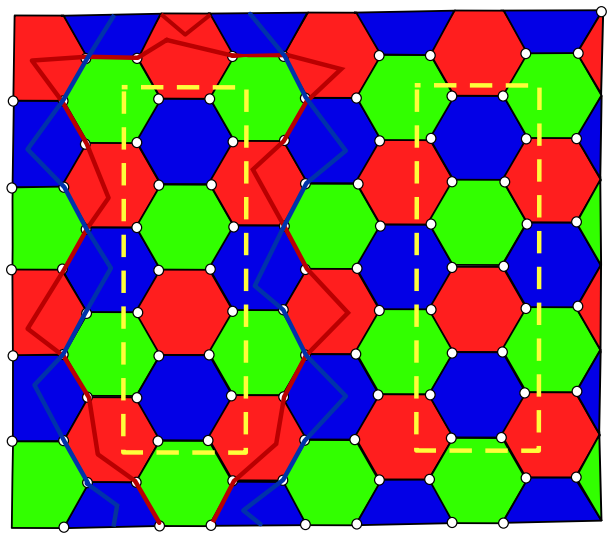} \caption{(color online)
One of subsystems, say \emph{A}, composed of two disjoint regions
which introduce some nontrivial colored closed strings acting solely
on \emph{B}. The closed blue and red strings have been shown only
for one region of subsystem \emph{A}.} \label{fig5}
\end{center}
\end{figure}
%%%%%%%%%%%%%%%%%%%%%%%%%%%%%%%%%%%%%%%%%%%%%%%%%%%%%%%%%%%%%%%%%%%%%%%%%%%%%%%%%%%%%%%%%%%
\subsection{Topological Entanglement Entropy}

As we see from the Eq.(\ref{eq29}), in the entanglement entropy of a
region there exists a subleading term which is universal and
independent of the shape of region. This is a characteristic
signature of topological order which has been found on the
subleading term of the entanglement entropy. To be more precise on
the topological character of subleading term, we can construct a set
of bipartitions in which the boundary contribution is dropped and
remaining term is stemmed from the topological nature of the code
and inspires the topological order. To drop the bulk and boundary
degrees of freedom we consider a set of partitions which have been
shown in Fig.(\ref{fig4}). Topological entanglement entropy then
arises from the following combination of entanglement entropy \bea
\label{eq35}S_{topo}=\lim_{R,r \rightarrow
\infty}(-S_{1A}+S_{2A}+S_{3A}-S_{4A})~.\eea

For each bipartition one can use the entanglement entropy as before,
namely $S_{G}=\log(|G|)-\log(d_{A}d_{B})$. So far the cardinalities
$d_{A}$ and $d_{B}$ were determined by the number of plaquette
operators acting solely on \emph{A} and \emph{B}, respectively, i.e.
$d_{A}=2^{\Sigma_{A}}$ and $d_{B}=2^{\Sigma_{B}}$, where
$\Sigma_{B}$ and $\Sigma_{B}$ stand for the number of plaquette
operators acting only on \emph{A} and \emph{B}, respectively.
However, for calculating of the $S_{topo}$ we should take into
account other closed strings in addition to the above closed strings
acting on subsystems. For the case in which each partition is a
single connected region, for example bipartitions $(2)$ and $(3)$ in
Fig.(\ref{fig4}), the previous arguments work, but for disjoint
regions such as $(1)$ and $(4)$ in Fig.(\ref{fig4}) we should extend
the above result. As an example a simple case has been shown in
Fig.(\ref{fig5}) where we have supposed the subsystem \emph{A} is
composed of two disjoint regions (two dashed rectangles which we
label them as $\emph{A}_{1}$ and $\emph{A}_{2}$ ) while the
subsystem \emph{B} is a single connected one. By inspection we see
that the product of two set of plaquettes, say blue and green ones,
which act on $\emph{A}_{1}$ and the blue and green ones acting
simultaneously on $\emph{A}_{1}$ and \emph{B} result in a red string
which acts only on \emph{B}. The same scenario can be applied for
other choices of the plaquettes, e.g. red and green or red and blue
plaquettes, which yield blue and green string leaving the subsystem
$\emph{A}_{1}$. However, as we pointed out before one of them will
be
immaterial since there is an interplay between homology and color.\\
The main point is that it is not possible to produce these colored
closed strings which act only on \emph{B} from the product of some
plaquettes of region \emph{B}. The same thing will be held for the
another disjoint region of \emph{A}, say $\emph{A}_{2}$. So, we have
a collection of closed strings with nontrivial effect on \emph{B}
and we must take into account them in calculation of $d_{B}$. But, a
remark is in order and that if for example we combine two red
strings, the resultant is not a new closed string because we can
produce these two red string from the product of the blue and green
plaquettes of region \emph{B}. Therefore for the case shown in
Fig.(\ref{fig5}) there are only two independent closed strings of
this type with a nontrivial effect on \emph{B}. With these remarks,
now the cardinalities $d_{A}$ and $d_{B}$ are

\bea \label{eq36}
d_{A}=2^{\Sigma_{A}}~~~,~~~d_{B}=2^{\Sigma_{B}+4-2}~.\eea

We can generalize the above results for the case that each subsystem
is composed of several disconnected regions. Let the partitions
$\emph{A}$ and $\emph{B}$ are composed of $\emph{m}_{A}$ and
$\emph{m}_{B}$ disjoint regions, respectively. So, the cardinalities
will be

\bea \label{eq33}
d_{A}=2^{\Sigma_{A}+2\emph{m}_{B}-2}~~~,~~~d_{B}=2^{\Sigma_{B}+2\emph{m}_{A}-2}~.\eea

Now we move on in order to calculate the topological entanglement
entropy defined in Eq.(\ref{eq35}). For the partitions shown in
Fig.(\ref{fig4}) we have

\bea \no \emph{m}_{1B}=\emph{m}_{4A}=2 ,\eea

\bea \label{eq36}
\emph{m}_{1A}=\emph{m}_{2A}=\emph{m}_{2B}=\emph{m}_{3A}=\emph{m}_{3B}=\emph{m}_{4B}=1
,\eea

\bea \label{eq37}
\Sigma_{1A}+\Sigma_{4A}=\Sigma_{2A}+\Sigma_{3A}~.\eea

Finally the topological entanglement entropy reads as follows: \bea
\label{eq38} S_{topo}=-4~~.\eea

Topological entanglement entropy depend only on the topology of the
regions and no matter with their geometries \cite{kitaev}. This
derivation for topological entanglement entropy of color codes is
consistent with the subleading term of scaling of the entropy. The
topological entanglement entropy is related to the subleading term
of the entanglement entropy, i.e. $S_{topo}=-2\gamma$
~\cite{kitaev}. This leads to $\gamma=2$ which has also been derived
Eq.(\ref{eq29}).\\

%%%%%%%%%%%%%%%%%%%%%  Fig.6   %%%%%%%%%%%%%%%%%%%%%%%%%%%%%%%%%%%%%%%%%%%%%%%%%%%%%%%%%%%
\begin{figure}
\begin{center}
\includegraphics[width=8cm]{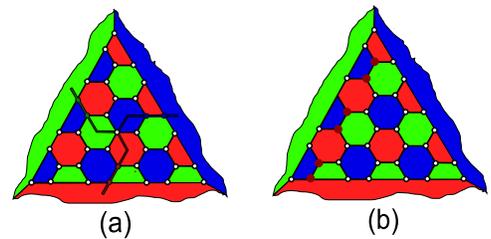}
\caption{(color online) A class of color code on the plane. It
includes borders each of one color. It encodes one qubits and can be
produced from a 2-colex without border by removing a couple of
plaquettes. To each border only strings with the same color of
border can have end points. (a) 3-string net commutes with all
plaquette operators. Such string-net and its deformation makes the
encoded Pauli operators acting on the encoded qubit. (b) A
manifestation of colored string as a bipartition. It stands for an
error.} \label{fig6}
\end{center}
\end{figure}
%%%%%%%%%%%%%%%%%%%%%%%%%%%%%%%%%%%%%%%%%%%%%%%%%%%%%%%%%%%%%%%%%%%%%%%%%%%%%%%%%%%%%%%%%%%

\section{entanglement properties of Planar color codes}

An important class of topological stabilizer codes in practice is
planar codes which are topological codes that can be embedded in a
piece of planar surface. These planar codes are very interesting for
topological quantum memory and quantum
computations\cite{dennis,bombin2}. A planar color code will be
obtained from a 2-colex without border when a couple of plaquettes
removed from the code. For example when a plaquette, say green, is
removed only green strings can have endpoint on the removed
plaquette not the blue and red strings. The same scenario holds for
other plaquettes. By this construction we will end with a 2-colex
that has three borders each of one color. To each border only
strings can have endpoint which have same color of that border. The
most important class of such planar color code is \emph{triangular
code} that has been shown in Fig.(\ref{fig6}). The essential
property of such code is determined via realizing a 3-string
operator and its deformation. We denote them by $T^{X}$ and $T^{Z}$
so that $\{T^{X},T^{Z}\}=0$ since they cross each other once at
strings with different colors. This latter anticommutation relation
can also be invoked by considering all qubits which makes the
triangular color code very interesting for full implementation of
the Clifford group without need for selective addressing
\cite{bombin2}. Instead of giving such fruitful properties, we are
interested in the entanglement
properties of the triangular code.\\
To have a concrete discussion, we consider a triangular lattice
which contains hexagons like what has been shown in
Fig.(\ref{fig6}). The total number of plaquettes and vertices are
$|P|=\frac{3}{2}k(k+1)$ and $|V|=3k(k+1)+1$, respectively, where $k$
in an integer number. The plaquette operators like those defined in
Eq.(\ref{eq1})have been used in order to define a stabilizer
subspace for this code. A main point about the planar code is that
there are no constraints on the action of plaquette operators, i.e
they are independent. For example if we product all red and green
plaquette operators, the resulted operator will not be same with the
product of red and blue plaquette operators, a property that was
absent for color code on compact surface. However, the product of
distinct plaquettes is in the stabilizer group. The coding space is
spanned with the states which are fixed points of all plaquette
operators. With the above identifications for the number of
plaquettes and vertices, the dimension of the coding space will be
$\Lambda=\frac{2^{|V|}}{2^{|P|}2^{|P|}}=2$. This implies that
triangular code encodes a single qubit. Two states are completely
determined by the elements of stabilizer group and 3-string
operator. Note that here 3-string operator plays a role like
nontrivial closed strings in the color code on the torus. For
triangular code the 3-string is nontrivial, i.e. it commutes with
all plaquette operators and is not product of some plaquettes. So
the stabilized sates will be

\bea \label{eq39} |0\rangle=|G|^{-1/2}\sum_{g\in
G}g|0\rangle^{\bigotimes |V|}~~,~~|1\rangle=T^{X}|0\rangle~~.\eea

A generic state can be in a superposition of the above states such
as $|\psi\rangle=\sum^{1}_{i=0}a_{i}|i\rangle$ where
$\sum^{1}_{i=0}|a_{i}|^{2}=1$. The reduced density matrix in
Eq.(\ref{eq11}) gives what we need to evaluate the entanglement
entropy. Again we make different bipartitions and calculate the
entanglement entropy.\\Let the system be in the state $|0\rangle$. A
single qubit is maximally entangled with rest of the system while
two qubits are not entangled implying that the triangular code is
also a multipartite entangled structure. As an another bipartition
we consider a colored spin chain such as red chain in
Fig.(\ref{fig6}b). This string contains $2k$ qubits and anticommutes
with plaquettes which the string has an endpoint on them, i.e they
share in odd number of qubits, thus its action on the coding space
amounts to error or on the other hand to the appearance of
excitations above the ground state if we adopt the coding space as
ground state of a local Hamiltonian as in Eq.(\ref{eq10}). However,
such a string has maximal entanglement $S_{A}=2k$ with other qubits
of the lattice. This means that all configuration of the spin chain
are allowed in the state $|0\rangle$ of the code, a feature that is
similar to the entanglement of an open string in the topological
color code on the torus (see Subsec(C)).\\Taking into account the
qubits living on the 3-string net as subsystem $A$ leads to
occurrence of configurations of subsystem $A$ with an only even
number of spin flipped in the state $|0\rangle$. For this string net
as shown in Fig.(\ref{fig6}a) the number of qubits of the string is
$2k+1$, and the entanglement entropy then will become $S_{A}=2k$. By
inspection we see that the plaquette operators flip only even number
of qubits of the string net. Now let the system be in a generic
state such as $|\psi\rangle$. All possible configurations of string
net are allowed since all of them can be realized by applying some
3-string operators $T^{X}$. There are $2^{2k}$ configurations with
even number of spins flipped and $2^{2k}$ configurations with odd
number of spins flipped. The states $|0\rangle$ and
$|1\rangle$include even and odd configurations, respectively.
Therefore the entanglement entropy for a generic state will be:
$S_{A}=2k+H(\alpha)$ where $\alpha=|a_{0}|^{2}$.\\ What about the
topological entanglement entropy of planar color codes? In order to
answer this question we obtain a set of bipartitions such as in
Fig.(\ref{fig4}) in a large triangular code. Let $\Sigma_{A}$ and
$\Sigma_{B}$
 stand for the number of plaquette operators acting only on
$A$ and $B$, respectively, and $\Sigma_{AB}$ stands for the number
of plaquette operators acting simultaneously on $A$ and $B$. All
plaquette operators are independent. To get things simpler, let
consider a simple case where the subsystem $A$ is a single  convex
connected region. The cardinality of the subgroup $G_{A}$ is
$2^{\Sigma_{A}}$. However, there are two non-trivial closed string
acting on subsystem $B$ which is impossible to provide them by
product of some plaquettes of $B$. In fact these  two independent
strings are resulted from the product of some plaquettes, say red
and green, which are free to act on $A$. So the cardinality of the
subgroup $B$ reads $2^{\Sigma_{B}+2}$. Thus for the entanglement
entropy we obtain $S_{A}=\Sigma_{AB}-2$.\\In order to calculate the
topological entanglement entropy, first  the entanglement entropy is
calculated for different bipartitions  and then the expression in
Eq.(\ref{eq35}) gives the topological entropy. Because of the
boundaries in the planar codes, for a disconnected region like (1)
or (4) in Fig.(\ref{fig4}) the constraint we imposed in
Eq.(\ref{eq33}) is no longer true. For the bipartition (1) in
Fig.(\ref{fig4}) the cardinalities of subgroups $G_{A}$ and $G_{B}$
will be $d_{A}=2^{\Sigma_{1A}+2}$ and $d_{B}=2^{\Sigma_{1B}+2}$,
respectively. However, for the bipartitions (4) in Fig.(\ref{fig4})
the cardinalities will be $d_{A}=2^{\Sigma_{4A}}$ and
$d_{B}=2^{\Sigma_{4B}+4}$, respectively. Thus the topological
entanglement entropy becomes $S^{t}_{topo}=-4$. Although the closed
string structures of bipartitions in the triangular code due to the
boundary effects differ from the color code on the torus, the
topological entanglement entropy is same for both structures that is
related to the fact that both structures have the same symmetry.\\\

\section{conclusions}
In this paper we calculated the entanglement properties of
topological color codes (TCC) defined on a two-dimensional lattice.
The entanglement entropy was measured by the von Neumann entropy. We
considered two structures of TCC either defined on the compact or
planar surface. The coding space of TCC is spanned by a set of
states in which are the fixed points of a set of commuting Pauli
operators, i.e they are common eigenvectors of all elements of
stabilizer group with eigenvalue $+1$. In fact this set stabilizes
the coding space. The stabilizer group is generated from plaquette
operators. Product of different plaquette operators produces colored
strings which in fact are the closed boundary of several plaquettes.
However, There exist some independent non-trivial strings winding
the handles of manifold where the model defined on. This non-trivial
strings commute with all plaquette operators but are not actually in
the stabilizer group. These nontrivial strings make remarkable
properties of the color code more pronounced.\\ For a manifold with
genus $\mathrm{g}$, the coding space spanned by $4^{2\mathrm{g}}$
states that can be adopted in which be ground state of a local
Hamiltonian. The degeneracy of ground state subspace depends on the
genus of the manifold which is a feature of topological order.
Different states of ground state subspace can be constructed by
means of spin flipped elements of stabilizer group and nontrivial
closed loops. In order to calculate the entanglement properties of
the TCC, the reduced density matrix of a subsystem is obtained by
integrating out the remanding degrees of freedom. We did this by
using the group properties of the stabilizer group.\\ For both
structures of color code on compact and planar surface while a
single qubit is maximally entangled with others, two qubits are no
longer entangled. This finding manifests the color code is a
genuinely multipartite entangled structure. We also considered other
bipartitions such as spin chains, spin ladders with various colors
and homologies. For all bipartitions we found the entanglement
entropy depends on the degrees of freedom living on the boundary of
two subsystems. However, in the entanglement entropy of a convex
region, there is a subleading term which is ascribed to the
topological properties of region not the geometry, a feature that
has also been arisen in a massive topological theory\cite{kitaev}.
The fact that the entanglement entropy for a region of lattice
scales with its boundary is a feature of \emph{area law} which is
also of great interest in other branch of physics such as black
holes\cite{ryo}.\\We exploited the scaling of the entropy to
construct a set of bipartitions in order to calculate the
topological entanglement entropy of color codes. We find that it is
twice than the topological entropy in the topological toric code.
The non-vanishing value for topological entanglement entropy and
dependency of degeneracy on the genus of the surface demonstrate
remarkable features on the fact that topological color code maybe
fabric to show topological order. For the toric code model the total
quantum dimension of excitation is $4$ which stands for different
superselection sectors of its anyonic excitations. The total quantum
dimension of color code is $16$ inspiring that the topological color
codes may carry richer anyonic quasiparticles. They may be colored
and their braiding will have nontrivial effect on the wavefunction
of the system. Indeed braiding of a colored quasiparticle around the
another one with different color will give rise to global phase for
the wavefunction, i.e they are abelian excitations. Triangular color
code with a remarkable application for entanglement distillation is
also a highly entangled code. Although the colored strings employed
in the 2-colex in the compact surface are not relevant in the
triangular code, the notion of string-net gives some essential
properties to the triangular code. In the state of the triangular
code all configurations of a colored string, say a red one in
Fig.(\ref{fig6}b), are allowed, i.e is has maximal entanglement with
the rest of the system. However, for a string-net only the
configurations with even number of spin flipped are allowed.
Entanglement entropy scales with boundary of the region and incudes
a topological term. This topological terms yields a non-vanishing
topological entanglement entropy which is same with what we obtained
for the color codes on the torus. Similar entanglement properties of
color codes defined on the compact and planar surface are not a
matter of chance because both structures carry same symmetry.\\

%%%%%%%%%%%%%%%%%%%%%%%%%%%%%%%%%%%%%%%%%%%%%%%%%%%%%%%%%%%%%%%%%
I would like to thank A. Langari for fruitful discussions and
comments. I would also like to acknowledge M. A. Martin-Delgado for
his useful comments. This work was supported in part by the Center
of Excellence in Complex Systems and Condensed Matter
(www.cscm.ir).\\\\

%%%%%%%%%%%%%%%%%%%%%%%%%%%%%%%%%%%%%%%%%%%%%%%%%%%%%%%%%%%%%%%%

\end{document}